# Attention Dynamics and Adaptive Decision Support in C5ISR: A Recurrence Quantification Analysis of Visual and Multimodal Attention Guidance Effects on Mission Performance

Hyun-Gee Jei, Caleb J. Armstrong, and Farzan Sasangohar

Wm Michael Barnes Department of Industrial and Systems Engineering, Texas A&M University

Competing Interests Statement: The authors have no competing interests to report.

**Funding:** This research did not receive any specific grant from funding agencies in the public, commercial, or not-for-profit sectors.

**Author Contributions: Hyun-Gee Jei:** Conceptualization, Methodology, Validation, Formal Analysis, Investigation, Data Curation, Writing-Original Draft, Visualization, Project administration; **Mustafa Demir:** Formal Analysis, Data Curation, Writing-Original Draft, Visualization; **Farzan Sasangohar:** Conceptualization, Methodology, Resources, Writing-Reviewing&Editing, Supervision, Project administration
Manuscript type: Research Article

Exact word count: 7560

*Corresponding Author. Farzan Sasangohar, 101 Bizzell St, Mail Stop 3131, College Station, TX 77843-3131. Email: sasangohar@tamu.edu; phone: +1-(682)-300-8380

**ABSTRACT**

Modern command, control, communications, computers, cyber, intelligence, surveillance, and reconnaissance (C5ISR) environments place extraordinary attentional demands on mission commanders (MC). Failure to allocate attention properly in these high-risk, high-stakes environments can have severe consequences. This study investigates the efficacy of gaze-driven, attention-guided (visual-only and multimodal) adaptive decision support tools (DSTs) and the underlying gaze dynamics that may affect MCs' performance in a high-fidelity simulated military command center. To analyze gaze and attentional dynamics while interacting with the attention-guided adaptive DST, Recurrence Quantification Analysis (RQA) was applied to the eye-tracking data. Stepwise regression with BIC was also used to identify RQA metrics predictive of the performance score. Regression results revealed that multimodal adaptive DST predicted significantly higher performance than visual-only attention-guided adaptive DST. The RQA metric average diagonal line length (L) had a negative linear relationship with performance level, whereas entropy (ENTR) had a positive linear relationship. Additionally, recurrence rate (RR), determinism (DET), and ENTR exhibited nonlinear quadratic relationships with performance. More specifically, RR and DET followed an inverted-U pattern consistent with the Yerkes-Dodson law. These findings suggest that dynamic contexts require both structured and dynamic scanning behavior to perform best.

# 1. INTRODUCTION

Modern command, control, communications, computers, cyber, intelligence, surveillance, and reconnaissance (C5ISR) environments have become saturated with data, placing unprecedented demands on the cognitive limits of mission commanders (MCs). While advancements in AI and real-time sensing were intended to clarify and ease understanding of the battlefield, the sheer velocity and volume of incoming data often exceed human information-processing capacity. Maintaining situational awareness (SA) in these settings is no longer merely a matter of data access but a challenge of efficient visual attention allocation and mission success (Demir et al., 2023; Endsley, 2001; Endsley et al., 2003; Humr et al., 2023). When MCs must navigate high-volume interfaces, the risk of "information missed" becomes a critical vulnerability in the decision-making loop (Curts & Campbell, 2001; Endsley & Garland, 2000).

Information overload can create attentional bottlenecks in modern military command (Giles, 2019; Shackelford & Lily, 2026). It is well documented that human cognitive resources are finite and easily depleted (Rogers, 2023; Tsotsos et al., 1995) and that attention is naturally divided when information overload occurs (C. D. Wickens, 1976). In high-stakes contexts such as C5ISR, divided attention can lead to an operational bottleneck. While selective attention allows MCs to prioritize high-value information, it may lead to tunnel vision, causing potential misses of critical peripheral signals, significant response delays, or loss of SA (Endsley et al., 2003). In C5ISR, MCs' ability to manage attention is tested by the requirement to monitor multiple visual displays simultaneously. In such dynamic environments, precise gaze and conscious engagement, enabled by proper attention allocation, are necessary conditions for an effective information-processing loop that culminates in timely decisions. When the demand for divided attention exceeds the MCs' cognitive capacity, the resulting cognitive friction leads to failures in which vital data is perceived but not processed into a decision. Hence, today's battlefield and C5ISR environments demand aids to guide human commanders' attention.

Eye-tracking has been the most common method to investigate attention allocation in the human factors literature. Most eye-tracking studies analyze time-collapsed metrics such as fixation counts and durations within static areas of interest (AOIs). However, this common approach could miss the inherent temporal dependencies and dynamical organization of human gaze behavior (Mahanama et al., 2022). By incorporating the temporal dimension, eye-tracking analysis can reveal more nuanced insights into visual attention dynamics, such as scanpath structure, gaze (in)stability, and recurrent gaze patterns. Therefore, visual attention dynamics could serve as a differentiating factor across experimental conditions and as reliable predictors of overall decision-making performance.

In a previous study by Jei et al. (2026), the authors tested the adaptive attention-guided decision support tool (DST) for MCs overseeing a ground convoy mission. In the experiment, the adaptive attention-guided DST significantly improved participants' overall performance. This improvement was amplified when participants' experience was combined with the adaptive DST. However, one caveat of the previous study was that the adaptive attention guidance was provided only in a visual format. In environments where commanders must gather information from multiple displays, some displays may be completely out of the commander's field of view. In that case, no matter how salient the attention guidance is, that information will be missed. Therefore, a need for a multimodal adaptive attention-guided DST was raised. According to Multiple Resources Theory (MRT; Wickens, 2008, 2021), human cognitive resources are depleted more quickly when information is presented through the same channel (i.e., visual) because multiple pieces of information must compete for the same cognitive resources. However, information is less likely to compete for cognitive resources when presented across different channels (e.g., visual and auditory). Additionally, when information is presented in different forms, its eccentricity and saliency also increase (C. Wickens, 2008, 2021). According to the N-SEEV model, the probability that a human operator notices an event or piece of information increases with eccentricity and saliency (Wickens, 2015; Wickens et al., 2009). The event's expectations and value also contribute to the increase in probability; however, these are top-down elements that are difficult to control through external nudges, such as attention guidance.

Several studies have found that combining modalities more effectively guides human attention to desired locations. For example, Reyes & Alles (2021) found that the combination of visual and auditory cues improved attention guidance in a cyber-physical environment. Moreover, Sheth & Shimojo (2004) and Van der Burg et al. (2009; 2008; 2008) showed that multimodal stimuli boosted the saliency of visual targets in visual search tasks. Although multimodal attention guidance has been deployed in automotive (Calvi et al., 2021) and military decision-making contexts (Savick et al., 2008) and has shown promising results in improving human performance, to our knowledge, investigating multimodal real-time attention guidance in a military C5ISR supervisory control context remains an open gap. To address this gap, this study documents an experiment conducted in a simulated C5ISR environment to assess 1) whether multimodal adaptive attention guided DST further improves MC's overall performance, 2) how visual attention dynamics differ across different conditions, and 3) what reliable eye-tracking measures can predict MC's performance.

## 2. METHODS

### 2.1. Study Design

A controlled experiment was designed following a mixed 2 (Condition) x 2 (Scenario) x 2 (Order) factorial design. The two within-subjects factors were Condition (Visual-only (V) vs. Visual and Auditory (VA)) and Scenario (Scenario 1 vs. Scenario 2), with the order of the two factors counterbalanced between-subjects. With this experimental design, the main effects of Condition and Scenario were observed, as was the potential learning effect of exposure order. All participants completed two trials: one under the V condition, in which the adaptive DST was presented only with visual cues, and another under the VA condition, in which the adaptive DST was presented with both visual and auditory cues to guide participants' attention. The two scenarios were designed to have equivalent difficulty by having the same number of threats and resources.

### 2.2. Sample Size and Participants

A power analysis was conducted using G*Power 3.1 (Faul et al., 2007) to obtain an appropriate sample size. The analysis was done using an F-test, with a medium effect size ($\eta_p^2$ = 0.06; Cohen, 1988), and an alpha ($\alpha$) of 0.05. For the mixed design study, the required sample size was 34 to achieve a power of 0.80. Thirty-five participants from the Corps of Cadets and the Veteran Student Association (VSA) were recruited at a large Southwestern University in the U.S. This population was used to leverage existing military training and educational experiences (Fletcher, 2009). The inclusion criteria required participants to be 18 years or older, be fluent in English, be comfortable using a standard computer mouse and keyboard, and have normal or corrected-to-normal hearing and vision, which was self-reported (e.g., colorblind participants were excluded). This study was approved by the university's Institutional Review Board (IRB), and each participant was compensated $50 for the 3 hours spent completing the experimental sessions.

### 2.3. Simulation Testbed

The study was conducted in a simulated environment that used a C5ISR task in which the MC oversaw a ground force protection mission, using various assets and resources to safeguard a critical political convoy as it traveled through a hostile region. The participants assumed the role of an MC and teamed up with three computer-simulated Unmanned Aerial Vehicle (UAV) operators, each controlling three semi-autonomous UAVs. These operators used UAVs to scan designated areas to discover the hostile force's missiles of various ranges in their AOI. The friendly force also coordinated with an external air-strike team (simulated) to destroy any enemy threats identified. The resources available to the human MC were UAVs, an airstrike team, and E-3 Airborne Warning and Control Systems (E-3 AWACS).

The simulation testbed comprised three 70-inch touchscreen displays (Dell C7017T) and a 55-inch interactive tabletop display (**Error! Reference source not found.**). To emulate real-world command centers, each display presented different types of information to MCs, enabling them to complete the mission successfully. First, the Mission Status Display (MSD) showed the UAV operators' performance metrics, the UAVs' status, and the communication link status. The MSD also showed important information on the UAVs' activity status so the MC could request threat information and identify the threat type manually (e.g., short-, medium-, or long-range missiles) if the UAVs did not do so within 30 seconds. The information requested by the MC to manually identify threats was presented on the Remote Assistance Display (RAD). The Map Display (MD) was the main display on which the MC could track the convoy's current location and status, areas scanned by the UAVs, discovered threats and their ranges, and the elapsed mission time. The MD also included the timeline, which showed the strike schedules, potential and known threat envelop, the timing of UAV neutralization, and the convoy attack. All the decisions and orders issued by the MC were communicated to other assets via the tabletop Mission Commander Display (MCD). As the MCs oversaw and tracked the mission progress, they were required to make decisions to keep the convoy safe. The types of decisions available on MCD were stopping/releasing the convoy, reporting any late strikes by the airstrike team, rerouting the UAVs to different areas, and changing the convoy's path to a safer route. The mission's success depended not only on monitoring its status but also on the MCs' ability to make timely decisions to ensure the convoy's safety.

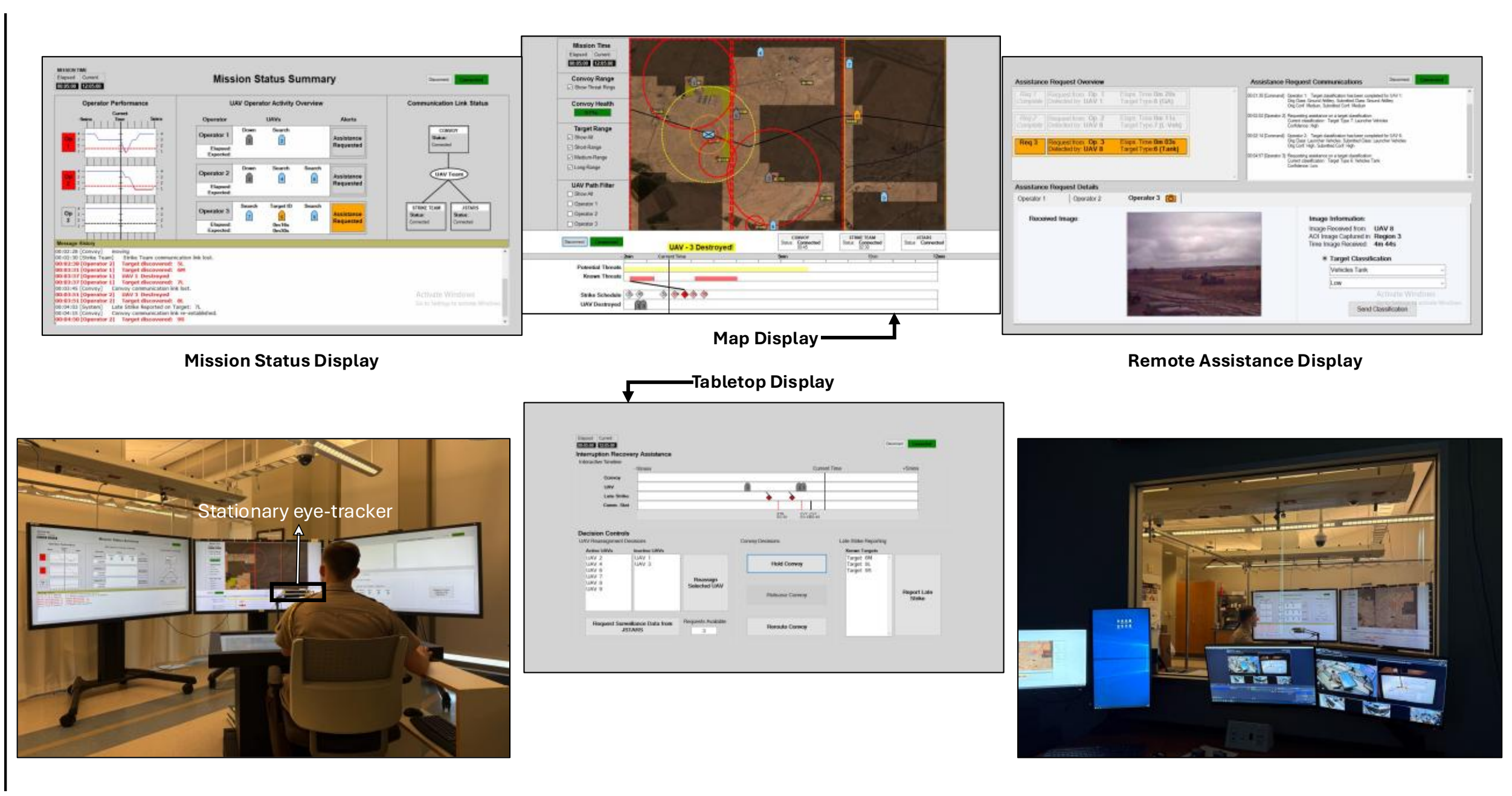

**Figure 1**. Simulation testbed and its display layout. Mission commanders monitor and make decisions based on the information gathered via simulated UAVs and E-3 AWACS.

## 2.4. Eye-Tracker-Based Adaptive Multimodal DST

The intervention in this experiment was based on real-time gaze data, which determined when attention-guiding elements were presented to participants. To enable the intervention, a stationary eye tracker, the Tobii Pro Nano, with a sampling rate of 60Hz and an accuracy of 0.3, was used. The eye-tracker was installed in front of the main map display to monitor the MCs' gaze behavior.

The eye-tracker's real-time gaze data was incorporated to support the adaptive attention-guided DST. The simulation's backend tracked events and where gazes landed. If the MCs failed to notice crucial events (e.g., a UAV being destroyed) while attending to other displays, the system would present attention-guidance to direct their attention to that event. In this experiment, attention guidance was provided in two formats: visual-only (V) and visual-auditory (VA). During the V session, various visual cues such as a salient warning message, color changes, and a blinking red box were used to guide the MCs' attention to the AOI. For example, when a friendly force's UAV gets destroyed by an enemy missile, a notification message would appear below the geographical map. Additionally, when the communication link between the MC, E-3 AWACS, and the airstrike team goes down, the communication status box would blink red. Lastly, a red box would also blink in the timeline to notify of any late strikes (**Error! Reference source not found.**). In the VA experiment session, auditory cues, such as a short beep and a verbal auditory message, were presented in addition to the aforementioned visual cues. However, since this is an 'adaptive' attention-guidance system, these visual and multimodal cues were given only when the MCs' visual attention was not in the right place at the right moments. If they were already looking at the critical events, no attention-guiding features were presented to prevent potential information overload and visual clutter. The location of the visual cues was chosen based on the proximity compatibility principle (C. D. Wickens & Carswell, 1995) and the N-SEEV model (C. Wickens, 2015). This placement is intended to minimize the effort required to notice and redirect attention to those cues.

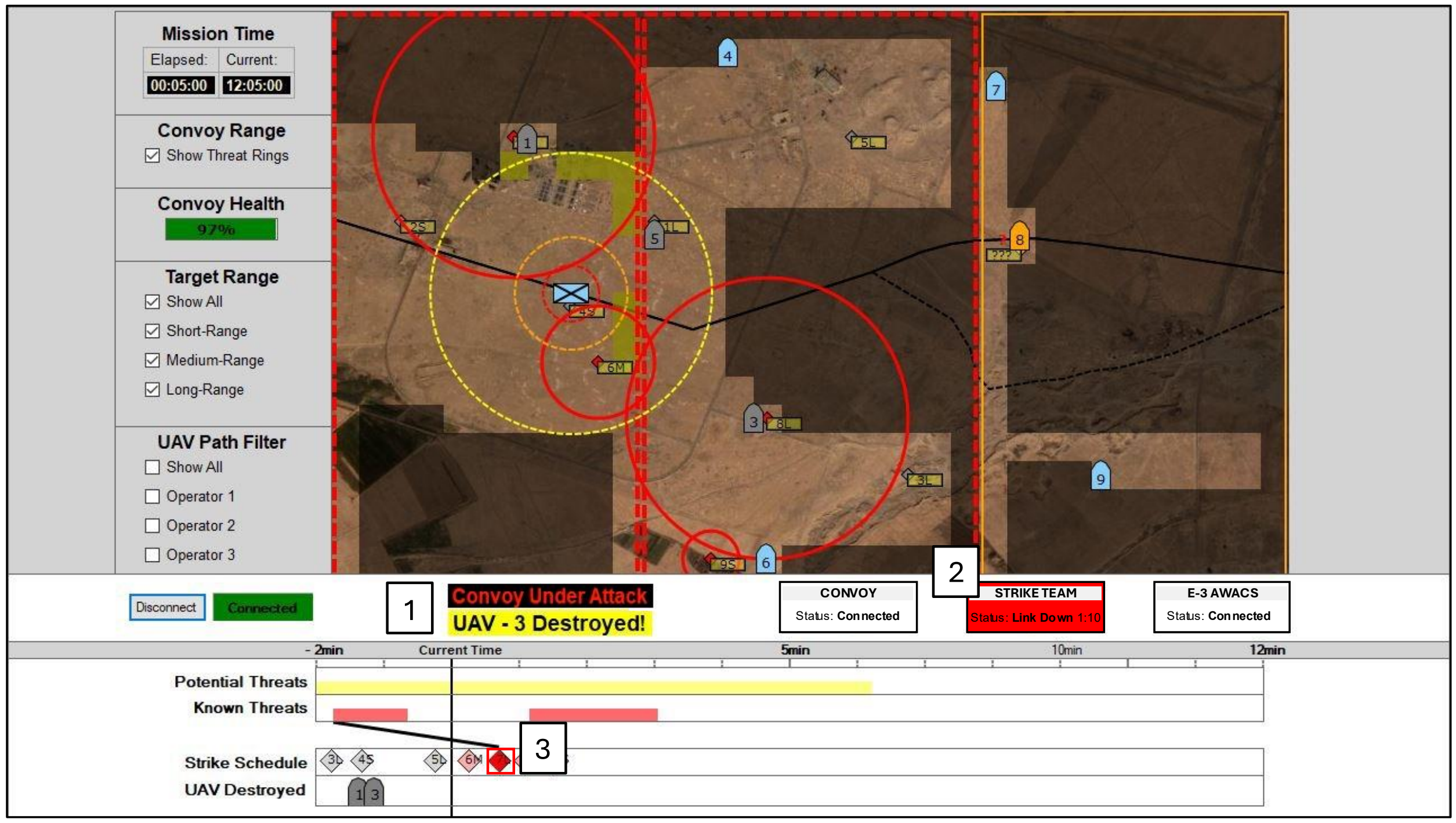

**Figure 2.** Example of the adaptive attention-guides displayed on the mission map display. Auditory sound is played simultaneously with the visual cues during the visual-auditory condition. Box 1: message below geographical map; Box 2: Red blinking communication link issue; Box 3: late strikes

## 2.5. Procedure and Training

As shown in Figure 3, the experiment consisted of a single 3-hour session, including a tutorial, benchmark testing, two trial sessions, and a post-experiment interview. Upon arrival, participants were briefed on the experiment and given time to review and sign the informed consent form. They were then given a PowerPoint tutorial about the experimental procedure, the mission summary, the mission goals, and descriptions of the functions and symbols shown on each display. At the end of the tutorial, six slides presented the different scenes that participants may encounter during the experimental trials. Experimenters reviewed these slides with the participants to facilitate an interactive discussion on the best course of action (COA) for each case. The discussion session ensured participants' understanding and enabled them to think critically and adopt a 'mission commander mode' mindset. Participants were also briefed on how their performance would be measured. The final training stage was to complete a training scenario on the actual experiment testbed. This stage also served as the benchmark test to ensure that the participants had sufficient knowledge of the mission and the testbed before convening the experimental trials. To move on to the experimental trials, participants were required to have at least 25% of the convoy's health remaining and to have utilized the critical functions and available reconnaissance assets. Participants had to repeat the training scenario if they did not meet the benchmark requirements. The tutorial and training session lasted 45-60 minutes, depending on participants' reading and comprehension speeds.

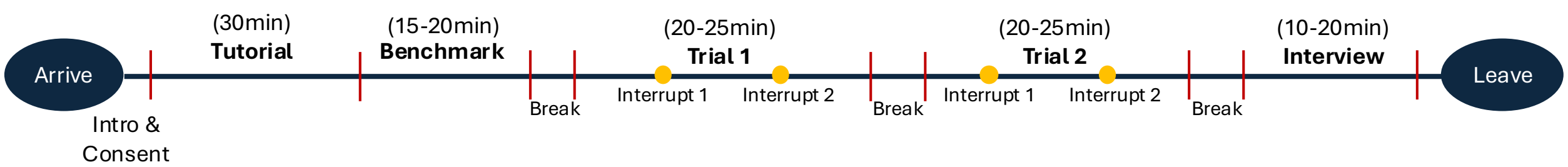

**Figure 3.** Timeline of the experiment.

Participants then completed two experimental trials. The order of the conditions and scenarios was counterbalanced to prevent bias and minimize learning effects. After completing the first trial, participants were asked to subjectively rate their situational awareness (SA) using the Situation Awareness Rating Technique (SART; Taylor, 2011), then took a 5-minute break before beginning the second trial. The same SART assessment was given after the second trial to compare their SA between the two trials. During each trial, participants were also removed from the testbed twice to a location in the experimental lab where they could not see the simulation's progress. This interruption was given to emulate the frequent interruptions that MCs experience in real life. During the interruptions, they had to solve problems that were both cognitively and visually engaging (e.g., mission planning based on map features and finding a map location given geographical coordinates). These types of tasks were designed to shift their minds and focus away from the primary task. Participants were called back to the testbed when the convoy reached a designated location. Upon returning, they were all exposed to the same decision point, which required them to quickly regain SA and make operational decisions to keep the convoy safe. Overall, the two experimental trials took about 40-50 minutes.

The eye-tracker was calibrated before starting each trial using the nine-point calibration and validation procedure embedded within the Tobii Pro Lab software (Tobii AB, 2019). Lastly, participants were asked to complete a demographic survey and were interviewed to gather more information regarding their overall experience and decision-making process. The survey and interview lasted around 10-20 minutes.

### 2.6. Measures

During the experiment, we collected the following measures: Measure of Performance, eye-tracking metrics, SA rating, and interruption task scores. However, only the following measures were used to address the research questions.

*Measure of Performance (MOP).* For each trial, we used a scoring method that acknowledges the multifaceted nature of U.S. military operational success, which assesses the performance of friendly forces (Joint Staff J-7, 2011). The MOP, as a weighted composite score of each trial, is determined based on three critical elements: the convoy health remaining at the end of the trial (weighted at 50%), the trial completion time (35%), and the number of late strikes (15%); so, the MOP ranges between 0 to 100 (Jei et al., 2026). For each trial, $i$, convoy health remaining ($H_i$; range 0 -100), was converted to a normalized score $S_{H,i} = H_i/100$. Completion time ($t_i$) was normalized such that a shorter time means a better performance. First, the times of all participants in trial $i$ were ranked from 1 to 35. Then, the ranks ($r_i$) were converted to a normalized score, $S_{T,i} = 100 * ((35 - r_i)/34)$. Late strikes ($L_i$) were similarly normalized, $S_{L,i} = L_i/L_{max}$ where $L_i$ is the number of late strikes reported by the participants, and $L_{max}$ is the total number of late strikes observed during the trial. The final composite score was computed as $MOP_i = 100(0.50S_{H,i}+0.35S_{T,i}+0.15S_{L,i})$.

Furthermore, the scoring method and the priorities of each element were briefed to participants during the tutorial and training sessions. The highest-priority element was convoy health, as convoy protection was the mission's primary goal. At the beginning of the mission, the convoy started with 100% health, and its health progressively decreased with exposure to different enemy threats or if held in position for too long. Participants could track convoy health on the main map display. The second priority element was the completion time. Completion time was critical for convoy protection since the convoy was exposed to fewer threats as it spent less time on the road. Lastly, the number of late strikes reported was captured by the simulation testbed from the start to the end of each simulation run. Participants were required to report any late strikes that occurred after the convoy had arrived in the threat area and would have been attacked if the convoy had not been held before entering it.

*Eye Tracking.* In addition to using eye-tracking data to monitor participants' real-time visual attention allocation, several measures were collected for post-experimental analysis. The Tobii Pro Lab software automatically collected these measures during each experiment session. The collected data included the gaze locations (X and Y coordinates), total fixation duration (TDF), average fixation duration (ADF), number of fixations (NF), and average pupil diameter for both eyes (APD). The mentioned measures were counted only if participants' gaze landed in the specific AOIs shown in the annotated map in Figure 4.

**Figure 4.** Areas of interest (AOIs) on the map display. The display was divided into 21 AOIs to accurately capture the location of MC's gaze during the experiment.

*Visual Attention*. In addition to the analysis of typical eye-tracking data described above, an additional analysis was conducted to investigate the visual attention of MCs during the events and in the moments immediately after. This analysis would tell the rate of commanders noticing the major events when they first occurred (without the adaptive DST), what their action was after noticing the event (if any), the rate of DST being triggered, and whether the commanders shifted their focus to the event area after seeing the adaptive DST. This would allow us to determine whether our DST effectively helped commanders shift their attention to the most important information at that moment.

## 3. ANALYTICAL PERSPECTIVE

### 3.1. Non-linear Dynamical Systems (NDS)

Humans' behavior and physiological (re)actions often cannot be predicted using the traditional static input-output model. Such complex and adaptive behaviors can be analyzed using nonlinear dynamical systems (NDS). An NDS is a system in which a set of state variables continuously evolves over time through the interactions of its components, and the output is not directly proportional to the input, unlike linear systems (Thelen & Smith, 2007). For example, in an eye-tracking context, using an NDS framework may lead to much more inference about human visual attention and how it unfolds over time than can be captured by standard metrics such as fixation duration and saccade amplitude (Anderson et al., 2013). Visual attention in complex and dynamic tasks may involve continuously scanning multiple pieces of information appearing at different locations and adapting its scanning behavior accordingly, often in a quite sudden and nonlinear manner.

An NDS can also be understood as a trajectory in 'state space'. A state space is a multidimensional space in which every possible state of the system corresponds to a unique point; a random system uniformly and repeatedly fills this space within each state. In human systems, this kind of behavior allows humans to remain stable yet flexible enough to adapt to changes and challenges (Kiefer & Pincus, 2023). In C5ISR, MCs who browse multiple displays to gather information may develop scanning strategies over time to efficiently collect the information they need to make the best decisions. However, they would also need to adapt their strategies as different situations unfold. Analyzing eye-tracking data through an NDS lens could provide a deeper understanding of how their gaze behaviors and adaptive strategies changed over time (Anderson et al., 2013).

### 3.2. Recurrence Plots (RPs)

RPs are a NDS visualization method for identifying when a system returns to previously visited regions of its phase space, a multidimensional space. In categorical Recurrence Quantification Analysis (RQA), the RP indicates when specific discrete states or state sequences recur over time. A 21-category time series is pictured in Figure  to demonstrate how a categorical RP is built. There are three regimes separated by dashed lines: Regime A (between 1–200) cycles through categories 1–7, Regime B (between 201–400) exhibits long dwell times in categories 8–14, and Regime C (between 401–600) switches stochastically among categories 15–21 with intermittent bursts. The symbolic RP (top-right panel) is constructed using the equality rule $R_{ij} = 1(x_i = x_j)$, so a black point at (*i, j*) indicates that the system captures the same categorical state at times $i$ and $j$. The diagonal in Regime A reflects periodic revisiting of the same state sequence (recurrence at fixed lags). The block-like structures in Regime B arise from sustained occupancy (dwell) in particular categories, producing dense recurrences within each dwell interval. Regime C yields an irregular texture, consistent with less regular switching and bursty repetitions. The off-diagonal structures indicate similarities between segments when the same categories recur across different times.

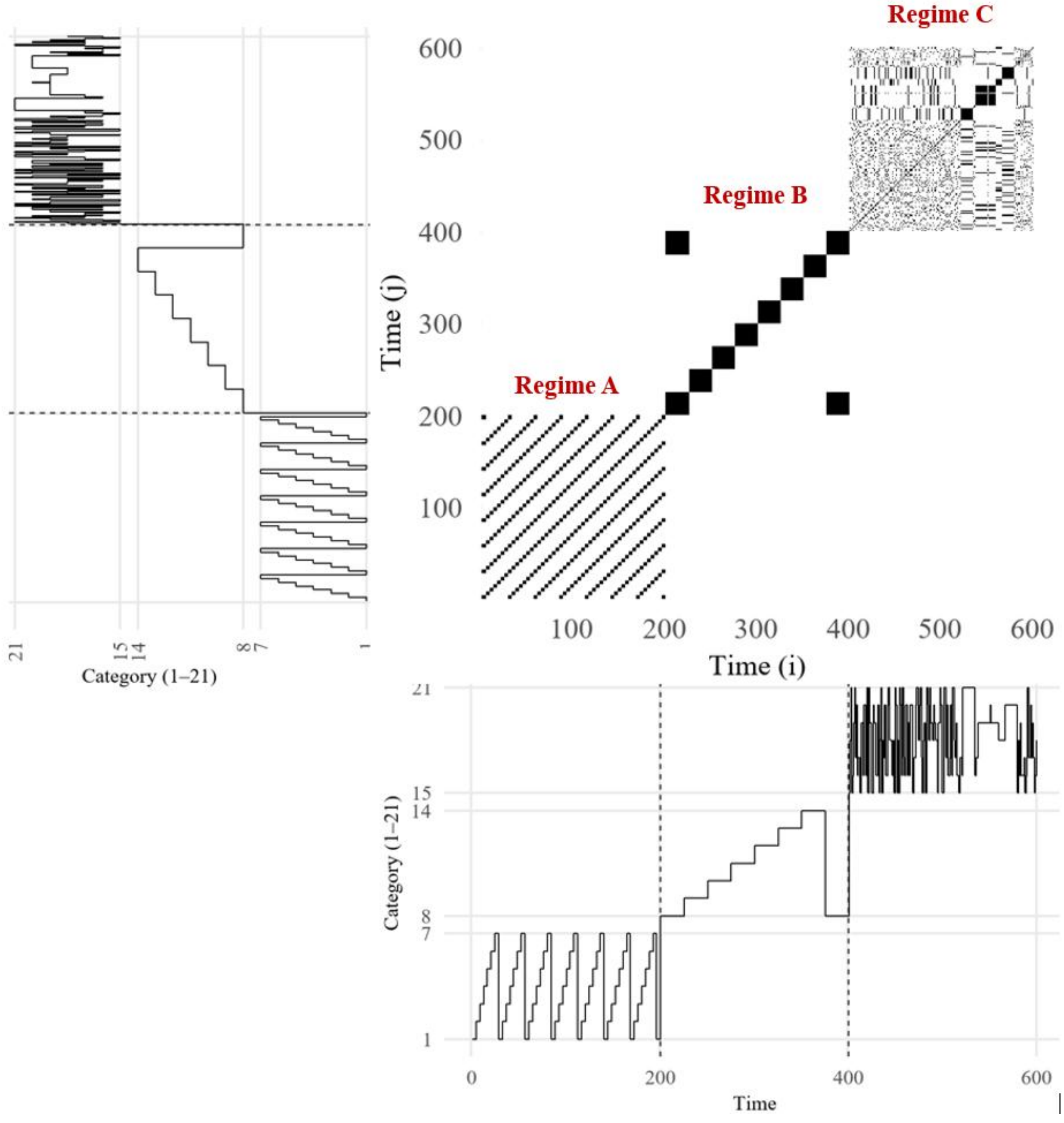

**Figure 5.** An illustration of how recurrence structure emerges in a discrete (categorical) time series. Because recurrence is defined as exact state matching, periodic cycling produces diagonal bands (Regime A), prolonged dwell times produce square blocks (Regime B), and irregular switching produces a more scattered texture (Regime C).

In the context of eye movements, the RP on the left in Figure  shows large, dense black boxes in the lower left corner. These boxes, created by recurrent points, indicate that the eye has returned to the same spatial location. Conversely, the RP on the right has more scattered recurrent points. As such, the visual patterns in the RP could reveal the hidden structure of the behavior.

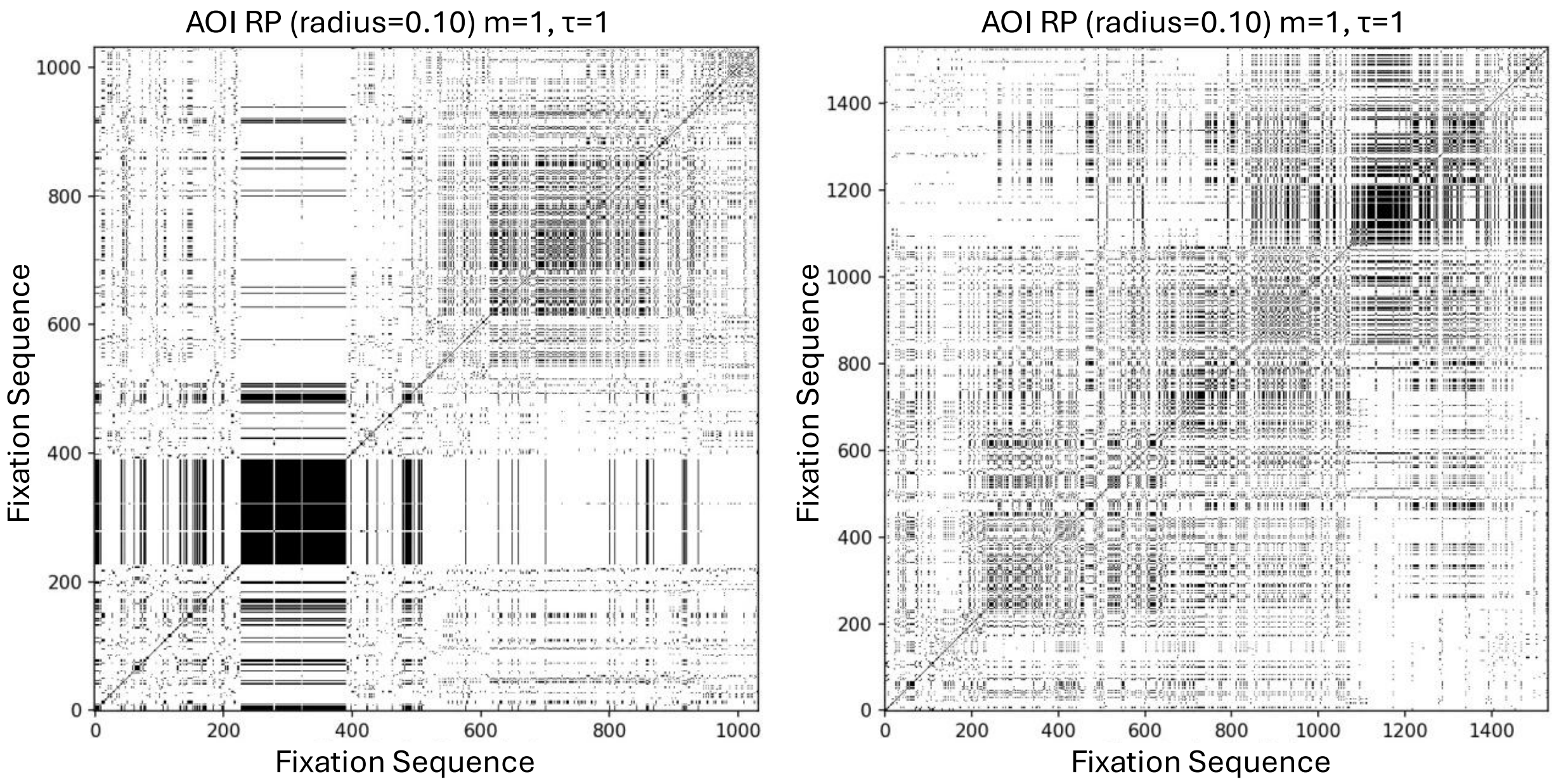

**Figure 6.** Example recurrence plots of repeated versus scattered gaze patterns.

### 3.3. Recurrence Quantification Analysis (RQA)

One of the NDS methods is RQA, which serves as the basis of RPs. RQA quantifies any moments when the system returns to a state it has visited before, as mentioned in the RP example (Zbilut & Webber, 1992). RQA has been successfully applied in various fields in terms of behavioral, such as team communication (Gorman et al., 2012)  and coordination (Demir et al., 2019, 2020), texting and reading (Allen et al., 2017; Likens et al., 2018), or physiological, such as eye-tracking (Atweh & Riggs, 2025) and heart rate (Wessel et al., 2001). This study examined whether the MCs' decision-making performance would be enhanced, depending on whether a V or a VA adaptive decision support tool was present. In terms of eye-tracking, one research question was how MCs' visual attention dynamics would differ across the two conditions, and the second was what eye-tracking measures could reliably predict their overall performance score.

In this study, each AOI was assigned a number (Figure 4), and a recurrent point was recorded if the mission commander's gaze returned to the same AOI during the simulation. Figure  shows examples of RPs of a good and a poor performer. The big black block on the left corner indicates that the mission commander was staying in or repeatedly coming back to the same AOI. On the other hand, the plot on the right shows a more sparse spread of points throughout the plot. This indicates that the gaze explored various parts of the display rather than focusing on one particular area.

### 3.4. RQA Metrics

RQA provides various metrics to detect hidden structures in a dynamical system. Among those, we focused on commonly used ones, including Recurrence Rate (RR), Determinism (DET), Average Diagonal Line Length (L), Laminarity (LAM), and Entropy (ENTR). These measures are the key dynamical properties most relevant to C5ISR attention management: overall stability/revisiting (RR), sequential organization/predictability (DET), persistence of routines (L), and complexity/flexibility of scanning structure (ENTR). Metrics such as laminarity (LAM) and related vertical-line measures primarily characterize state "stagnation" (extended pauses/laminar phases). Given the present study's emphasis on scanpath structure and performance-linked attentional control (rather than dwell/immobility dynamics), we prioritized diagonal-line metrics and recurrence density to maximize theoretical alignment and reduce redundancy among highly correlated RQA indicators.

**Table 1.** RQA metrics used in the study and their interpretation in C5ISR attention dynamics.

| Metric | Definition | Characterization of C5ISR Attention Dynamics |
|---|---|---|
| **RR** | Proportion of recurrent points in the RP, indicating the overall likelihood that the system revisits previously observed states. | **Question:** How often does the MC's visual attention return to previously observed AOI states (i.e., how frequently the same gaze-state patterns recur)?<br>**Attentional revisiting/stability.** Higher RR indicates operators repeatedly return to similar gaze states/AOI configurations (routine monitoring loops), whereas lower RR reflects less revisiting (more novel scanning). |
| **DET** | Proportion of recurrent points that form diagonal lines of the RP that reflect the predictability of the system, i.e., repeated trajectories. | **Question:** How often do recurrent gaze patterns of MC form predictable, repeated sequences (i.e., how consistently similar trajectories repeat over time)?<br>**Structured scanpath regularity.** Higher DET suggests more consistent, repeatable attention sequences (standardized cross-check routines); lower |
| | | DET suggests less organized, more opportunistic scanning. Extremely high DET can indicate rigidity (over-routinization). |
| **L** | Mean length of diagonal line segments or the RP, i.e., indexes persistence of repeated sequential patterns (how long similar trajectories in the system continue). | **Question:** How long do similar gaze trajectories continue once they emerge (i.e., how persistent are repeated attention sequences before switching)?<br>**Sustained, locked-in routines.** Larger *L* reflects longer uninterrupted attention sequences (extended dwell in a routine pattern), which can support exploitation but may reduce responsiveness to emergent cues if overly persistent. |
| **LAM** | Proportion of RPs forming vertical line structures in the overall RP; indexes the periods where the system remains in the same or slowly changing states (a.k.a. laminar states) | **Question**: How often does the MC's gaze enter dwelling episodes, and how long do these low-switching (stagnant) attention states persist?<br>**Attentional dwelling.** Higher LAM indicates longer or more frequent dwells within the same AOI, or minimal switching (e.g., prolonged inspection/ monitoring of a single display region); lower LAM indicates shorter dwells and more rapid transitions across AOIs. Very high LAM may suggest attentional tunneling under workload or salience. |
| **ENTR** | The Shannon entropy of the diagonal line-length distribution captures the heterogeneity/ complexity of the recurrent structure (variety of pattern lengths). | **Question:** How variable is the length and diversity of repeated gaze trajectories (i.e., how heterogeneous are the recurrent sequence patterns rather than uniform)?<br>**Adaptive complexity/flexibility.** Higher ENTR indicates a richer repertoire of attentional sequences (flexible switching and multi-scale monitoring), often beneficial under changing information demands; very low ENTR suggests overly uniform, stereotyped routines. |

**Note**. The table is created by using the following studies: Eckmann et al., 1987; Marwan et al., 2007; Zbilut & Webber, 1992

## 4. RESULTS

### 4.1. Stepwise Regression with Bayesian Information Criterion and Model Diagnosis

To identify a parsimonious set of RQA predictors of MOP across the conditions and trials, a stepwise regression method with both forward selection and backward elimination was applied (Draper & Smith, 1998; Miller, 1990). Model selection was guided by the Bayesian Information Criterion (BIC), which imposes a stronger penalty on model complexity and thereby minimizes the risk of overfitting (Schwarz, 1978). Using both forward selection and backward elimination iteratively evaluates candidate predictors by adding and removing terms based on the BIC, which balances goodness-of-fit with model complexity.

The initial model included the main effects of condition and trial, as well as standardized linear ($z_{RR}, z_{DET}, z_L$, $z_{LAM}$ and $z_{ENTR}$) and quadratic terms ($z_{RR}^2$, $z_{DET}^2$, $z_L^2$, $z_{LAM}^2$, and $z_{\mathrm{ENTR}}^2$) for RQA measures and their interactions with condition and trial. So, the bidirectional procedure begins with an initial model and, at each step, considers adding predictors that improve model fit as well as removing predictors that no longer contribute once other terms are included (Hocking, 1976). This process continues until no further additions or removals yield an improvement in BIC, resulting in a parsimonious model that retains only predictors that provide independent explanatory value (Draper & Smith, 1998). All analyses were conducted in R software (R Core Team, 2026) within RStudio (Posit Team, 2025), utilizing the MASS package for stepwise model selection (Ripley et al., 2023), and standardized regression coefficients were calculated using the lm.beta package (Behrendt, 2014). Marginal effects and visualizations were generated using ggeffects and ggplot2 (Lüdecke, 2018; Wickham, 2010). To assess the robustness of the findings, we also fit a linear mixed-effects model with a random intercept for participant by using lme4 and lmerTest (Bates et al., 2024; Kuznetsova et al., 2016).

The final regression model accounted for 44% of the variance in MOP, $R^2 = 0.44$, with an adjusted $R^2$of 0.35, and demonstrated a significant overall fit, $F(9{,}60) = 5.17, p < .001$. According to the regression diagnostics of the final model, a visual inspection of the residuals via a histogram, Figure (left), revealed an approximately symmetric distribution centered near zero, with no substantial skewness. The normal Q–Q plot Figure (right) further indicated that the majority of residuals aligned closely with the theoretical normal line, aside from mild deviations in the lower tail, which corresponded to a limited number of negative residuals. These minor tail deviations are frequently observed in behavioral and performance data and are not indicative of serious violations of the normality assumption; consistent with these observations, the Shapiro–Wilk test did not reveal a statistically significant departure from normality, $W = 0.97$, $p = .072$. Given that the null hypothesis of normality was not rejected, both the visual diagnostics and the Shapiro–Wilk test show that the residuals from the BIC-selected model meet the normality assumption to a satisfactory degree. Therefore, the application of the linear regression framework is justified for the present analysis.

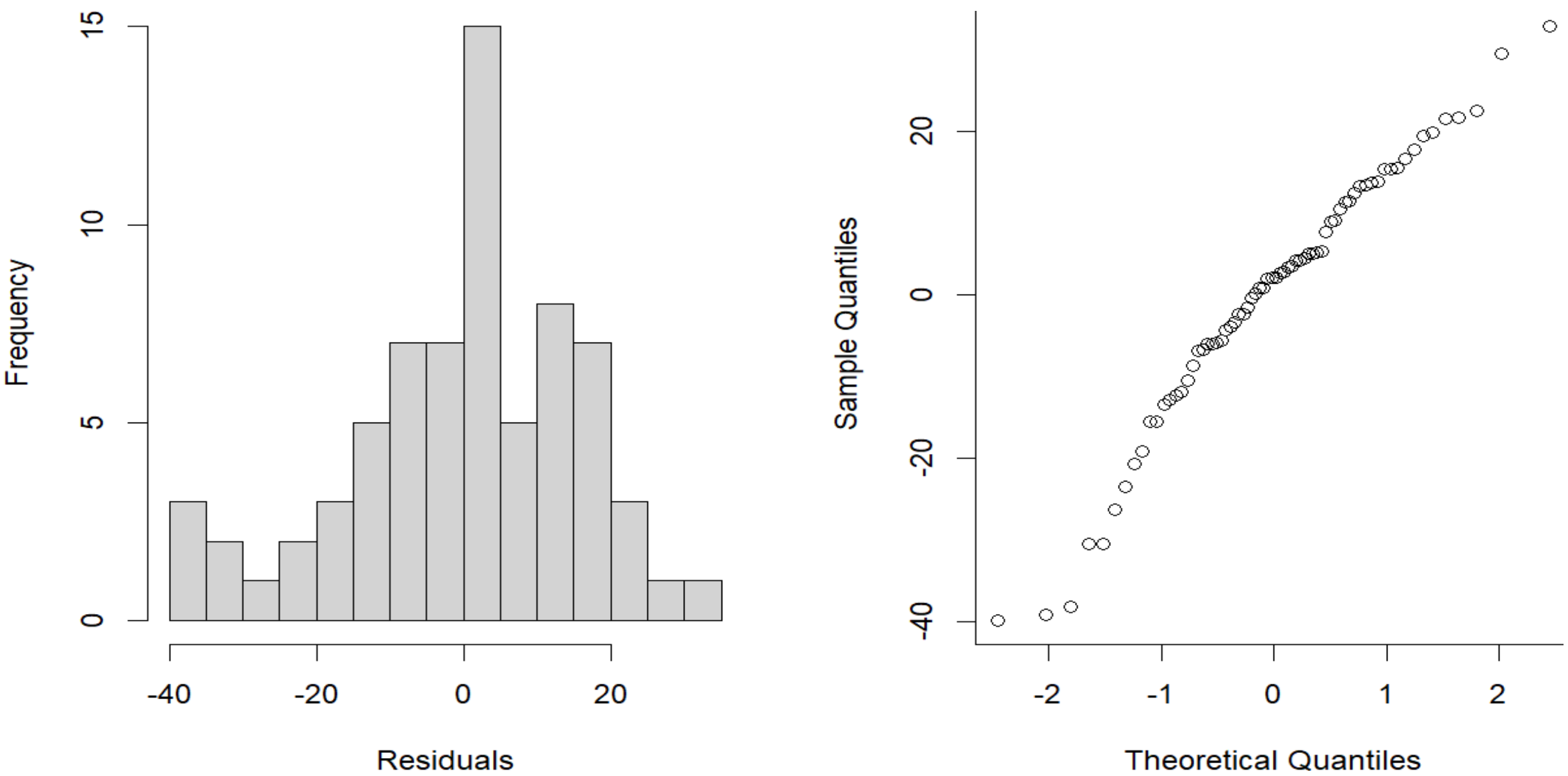

**Figure 7.** Residual diagnostics for the BIC-selected regression model predicting MOP: **(left)** The histogram shows an approximately symmetric distribution of residuals centered near zero, and (**right)** the normal Q–Q plot indicates close adherence to the theoretical normal distribution with minor tail deviations.

Formula 1 depicts the final model:

$$MOP_i = \beta_0 + \beta_1 Condition_i + \beta_2 Trial_i + \beta_3 z_{L,i} + \beta_4 z_{ENTR,i} + \beta_5 z^2_{RR,i} + \beta_6 z^2_{DET,i} + \beta_7 z^2_{ENTR,i} + \beta_8 (Trial_i \times z^2_{DET,i}) + \beta_9 (Trial_i \times z^2_{ENTR,i}) + \varepsilon_i. \quad (1)$$

Where $\beta_0$ represents the intercept, corresponding to the expected MOP score for the reference condition at Trial 1 when all standardized RQA predictors are at their mean values. Coefficients $\beta_1$and $\beta_2$represent the main effects of condition and trial, respectively. Coefficients $\beta_3$and $\beta_4$correspond to the linear effects of average diagonal line length ($z_L$) and entropy ($z_{\mathrm{ENTR}}$), capturing overall stability and complexity in interaction dynamics. Coefficients $\beta_5$, $\beta_6$, and $\beta_7$ represent the quadratic effects of recurrence rate ($z^2_{\mathrm{RR}}$), determinism ($z^2_{\mathrm{DET}}$), and entropy ($z^2_{\mathrm{ENTR}}$), respectively, thereby allowing nonlinear relationships between recurrence structure and MOP. Interaction terms $\beta_8$ and $\beta_9$ capture moderation by trial, indicating that the quadratic effects of determinism and entropy on MOP vary across trials. Finally, the error term $\varepsilon_i$ represents residual variance unexplained by the model.

The final regression results, summarized in Table 2 and formulated in Formula 2, revealed that the condition had a modest but statistically significant effect, indicating higher MOP scores in the VA condition relative to the reference condition (i.e., the V condition), as shown in Figure . Trial, however, did not show a significant main effect, suggesting that overall MOP levels did not differ between the two trials when other predictors were taken into account. Among the linear RQA metrics, the average diagonal line length ($z_L$) was negatively associated with MOP, whereas entropy ($z_{ENTR}$) showed a positive association ($p < .001$; see Figure ). These findings indicate that more stable, predictable interaction dynamics were associated with lower MOP scores, whereas greater temporal complexity was associated with higher MOP scores.

$$\widehat{MOP_i} = 44.762 + 8.835\, ConditionVA_i + 3.634\, Trial2_i - 26.440 z_{L,i} + 18.188\, z_{ENTR,i} - 3.300\, z^2_{RR,i} - 10.863\, z^2_{DET,i} + 9.319\, z^2_{ENTR,i} + 9.988\left(Trial2_i \times z^2_{DET,i}\right) - 11.014\left(Trial2_i \times z^2_{ENTR,i}\right) + \varepsilon_i. \quad (2)$$

**Note.** Condition $VA_i = 1$ for Visual & Auditory (0 = reference condition); $Trial2_i = 1$ for Trial 2 (0 = Trial 1). RQA predictors are standardized (z-scored).

**Table 2.** Standardized Regression Coefficients for the BIC-Selected Model Predicting MOP.

| Predictor | *b* | *SE(b)* | *β* | *t* | *p* |
|---|---|---|---|---|---|
| Condition (VA) | 8.84 | 4.40 | 0.21 | 2.01 | .049* |
| Trial 2 | 3.63 | 5.59 | 0.09 | 0.65 | .518 |
| Average diagonal line length ($z_L$) | -26.44 | 5.90 | -1.26 | -4.49 | <.001*** |
| Entropy ($z_{ENTR}$) | 18.19 | 4.19 | 0.86 | 4.34 | <.001*** |
| Recurrence Rate² ($z^2_{RR}$) | -3.30 | 1.36 | -0.28 | -2.42 | .019* |
| Determinism² ($z^2_{DET}$) | -10.86 | 3.53 | -0.67 | -3.08 | .003** |
| Entropy² ($z^2_{ENTR}$) | 9.32 | 2.82 | 0.71 | 3.30 | .002** |
| Trial x Determinism² ($z^2_{DET}$) | 9.99 | 4.29 | 0.59 | 2.33 | .023* |
| Trial x Entropy² ($z^2_{ENTR}$) | -11.01 | 4.37 | -0.52 | -2.52 | .014* |

**Note.** *b* = unstandardized coefficient, and its standard error *SE*(*b*); *β* = standardized coefficient. All RQA metrics were standardized prior to analysis. Trial 1 and Condition Visual Only served as reference categories. Model fit: $R^2 = .44$, Adjusted $R^2 = .35$, $F(9, 60) = 5.17$, $p < .001$. *$p < .05$. **$p < .01$. ***$p < .001$.

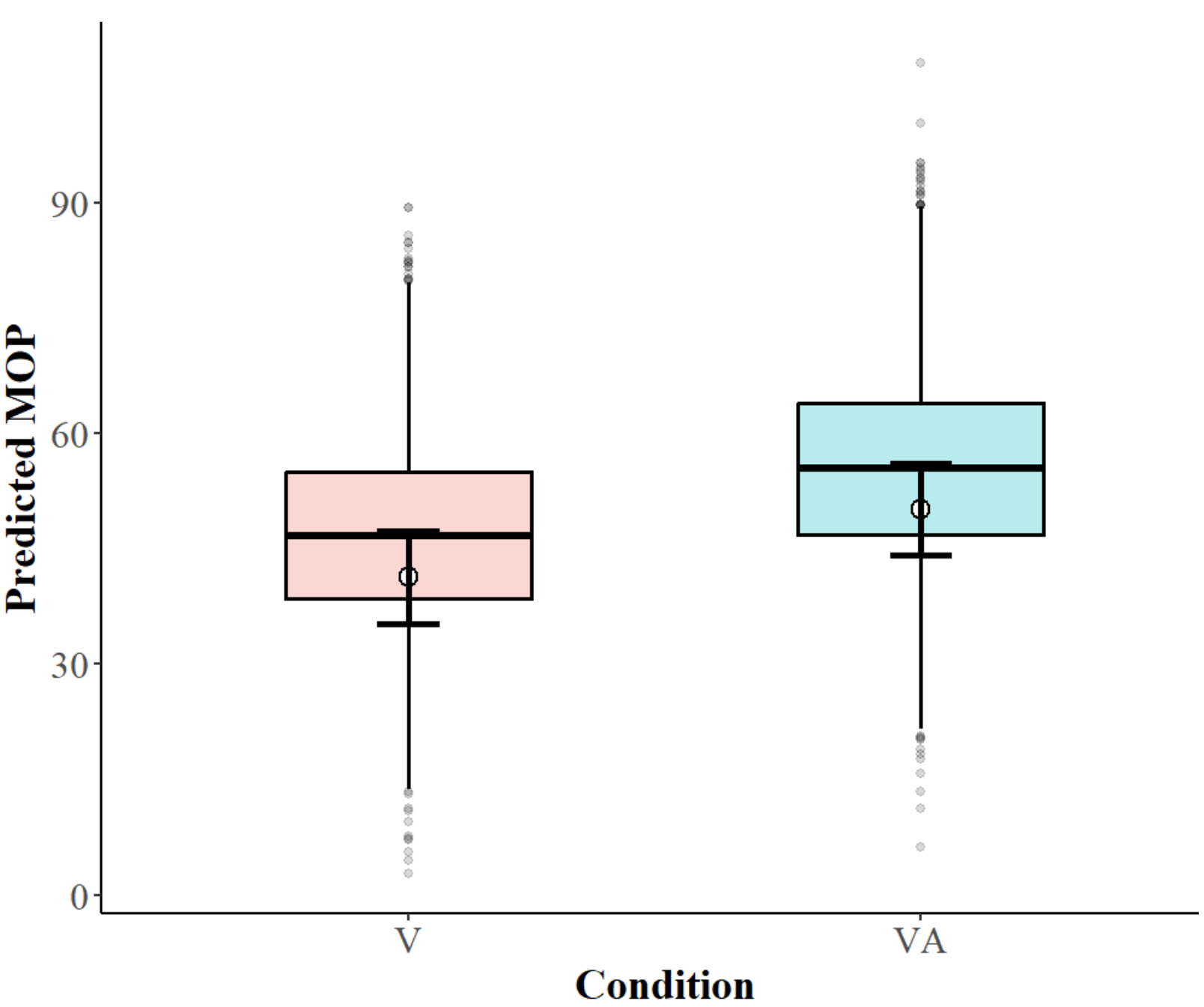

**Figure 8.** Model-based predicted MOP by condition (V vs. VA). Boxes summarize the model-implied predictive distribution of MOP for each condition: the box spans the interquartile range (IQR) with the median shown as the horizontal line; whiskers extend to 1.5 × IQR, and points denote outlying predicted values. White points indicate estimated marginal means (EMMs) from the fitted regression model (averaged over Trial), and vertical bars represent the corresponding 95% Confidence Intervals (CIs).

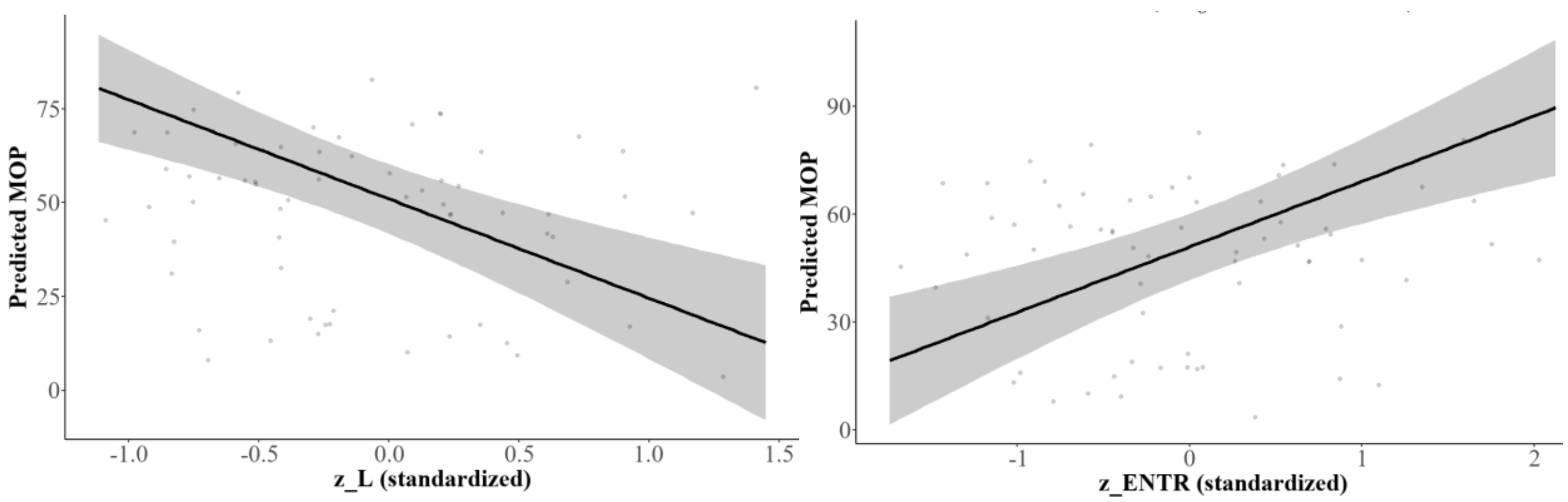

**Figure 9.** Model-based relationships between RQA metrics and MOP (averaged over Condition and Trial): Left panel: Predicted MOP as a function of average diagonal line length (z_L, standardized). Right panel: Predicted MOP as a function of entropy (z_ENTR, standardized). Solid black lines show the model-predicted mean MOP, with the shaded band indicating the 95% CIs for the mean prediction. Gray points represent observed MOP values overlaid for reference.

As seen in Table 2, the analysis also revealed significant nonlinear (quadratic) effects for recurrence rate ($z^2_{\mathrm{RR}}$), determinism ($z^2_{\mathrm{DET}}$), and entropy ($z^2_{\mathrm{ENTR}}$), respectively, indicating that the quadratic characteristics of the recurrence structure provided unique explanatory power for MOP, above and beyond linear effects; see Figure  for the direction of the relationships with MOP. Accordingly,  predicted MOP showed a quadratic inverted-U association with $z^2_{\mathrm{RR}}$ and $z^2_{\mathrm{DET}}$, with the highest MOP observed at moderate levels of each metric and lower MOP at both low and high levels. This nonlinear pattern suggests that moderate recurrence, meaning a balanced level of repeated structure in the attention dynamics, was associated with better performance, whereas both low recurrence (less structured dynamics) and very high recurrence (overly repetitive dynamics) were associated with lower MOP. For $z^2_{\mathrm{DET}}$, similarly, MOP was optimized when the attention dynamics demonstrated a moderate degree of sequential regularity (i.e., recurrence organized into coherent diagonal structures), whereas dynamics that were either weakly organized (low determinism) or overly constrained and predictable (high determinism) were associated with reduced MOP.

Predicted MOP increased with entropy ($z^2_{\mathrm{ENTR}}$), indicating higher MOP under more complex and heterogeneous temporal structure (Figure ); such that predicted MOP increased as $z^2_{\mathrm{ENTR}}$ increased, with the slope becoming steeper at higher entropy levels. This pattern implies that higher entropy, consistent with greater complexity and heterogeneity in the temporal organization of recurrent structures, was associated with higher MOP; as well as tended to be lower when attention dynamics were comparatively uniform (lower entropy) and higher when they reflected a richer, more varied distribution of recurrent patterns (higher entropy).

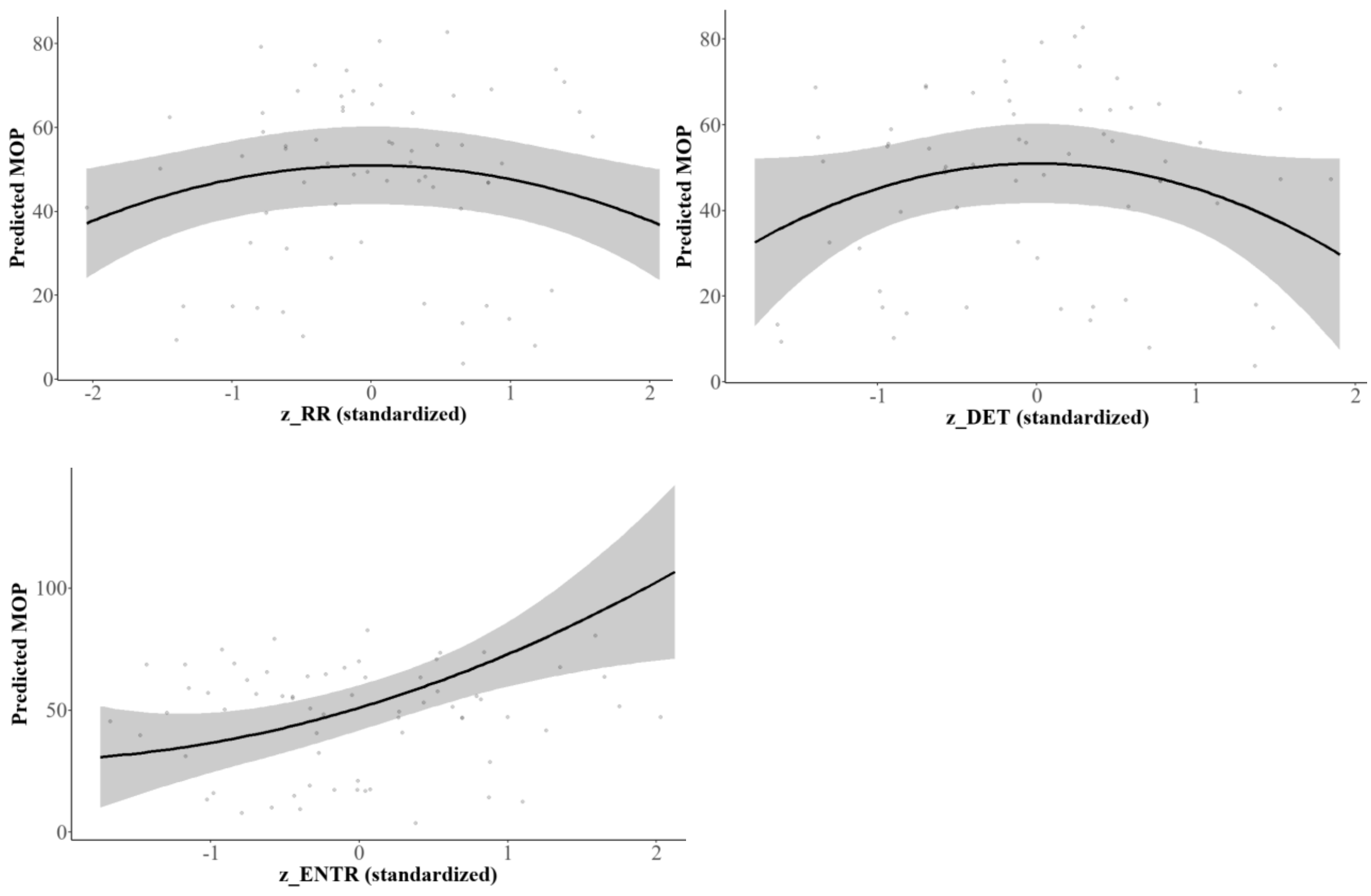

**Figure 10.** Model-based quadratic relationships between RQA metrics and MOP (averaged over Condition and Trial): Panels show predicted MOP as a function of (top left) recurrence rate (z_RR), (top right) determinism (z_DET), and (bottom) entropy (z_ENTR), each on a standardized scale. Curvature reflects the inclusion of the corresponding quadratic terms in the fitted model (i.e., $z^2_{\mathrm{RR}}$, $z^2_{\mathrm{DET}}$, $z^2_{\mathrm{ENTR}}$, respectively). Solid black lines indicate the model-predicted mean MOP; shaded bands denote 95% CIs for the mean prediction. Gray points are observed, MOP values overlaid for reference.

Notably, the quadratic effects of determinism and entropy were further moderated by trial, such that the relationship between MOP and these nonlinear RQA metrics varied with task exposure. These findings suggest that the influence of interaction structure on team performance is not static, but evolves with experience, reflecting both nonlinear and experience-dependent changes in the underlying dynamics. To probe these significant Trial × quadratic RQA interactions, specifically for $z^2_{\mathrm{DET}}$ and $z^2_{\mathrm{ENTR}}$, simple slopes (conditional effects) were computed for each trial and illustrated in Figure . For $z^2_{\mathrm{DET}}$, the association with MOP was significant at Trial 1, $b$ = -10.86, $SE$ = 3.53, $t$(60) = -3.08, $p$ = .003, a pronounced inverted-U association between $z^2_{\mathrm{DET}}$ and predicted MOP, consistent with optimal performance under moderately regular (but not overly rigid) scan sequences; yet, not at Trial 2, $b$ = -0.88,

*SE* = 2.87, *t*(60) = -0.30, *p* = .762. sequential regularity in attention dynamics no longer differentiated performance after additional exposure.

Likewise, for $z^2_{\text{ENTR}}$, the effect on MOP was significant at Trial 1, a strong positive nonlinear relationship, *b* = 9.32, *SE* = 2.82, *t*(60) = 3.30, *p* =.002, such that higher $z^2_{\text{ENTR}}$ reflecting greater heterogeneity and adaptive variability in recurrent attentional structure, was associated with higher MOP, whereas this relationship weakened by Trial 2, *b* = -1.69, *SE* = 3.42, *t*(60) = -0.50, *p* = .622. Overall, the results suggest that the nonlinear effects of determinism and entropy on the Measure of Performance (MOP) were apparent during the first exposure (Trial 1) but diminished by the second exposure (Trial 2). Overall, these interaction-effect findings indicate that the impact of nonlinear attention dynamics on team performance is strongest during initial exposure and diminishes as the operator adapts and stabilizes their attentional control strategies through repeated experiences.

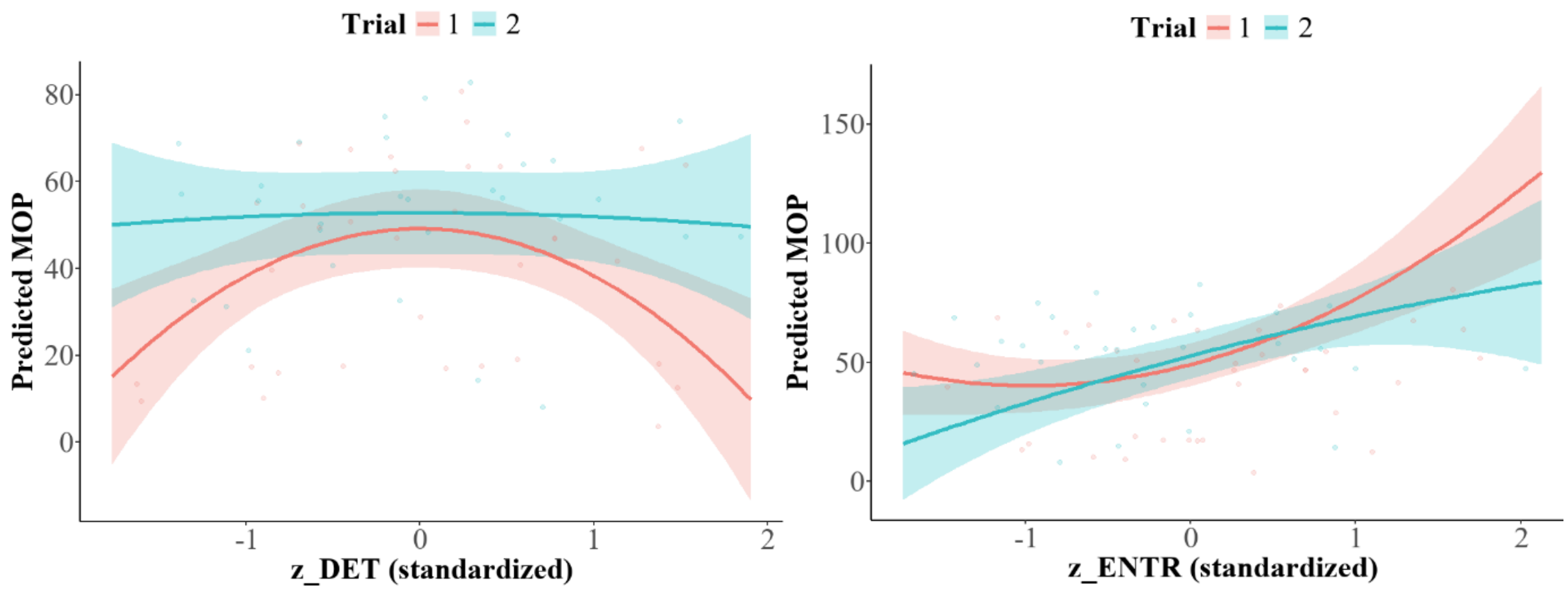

**Figure 11.** Panels depict model-predicted MOP as a function of (A) standardized determinism ($z^2_{\text{DET}}$) and (B) standardized entropy ($z^2_{\text{ENTR}}$), with separate curves for each Trial to visualize the significant Trial × $z^2_{\text{DET}}$ and Trial × $z^2_{\text{ENTR}}$, interaction terms. Solid lines represent model-predicted mean MOP; shaded bands indicate 95% CIs for the mean prediction. Points show observed MOP values overlaid for reference.

### 4.2. Visual Attention Dynamics Analysis

Due to the non-normal distribution of the data and factorial nature of the experimental design, an Aligned Rank Transform (ART) ANOVA was used to evaluate the differences in visual attention across conditions (see Table 3). Analysis revealed that the most significant visual attention difference between the two conditions was the rate at which MCs first noticed the major events without the adaptive DST. The ART ANOVA indicated a significant main effect of Condition on the "Notice the event without DST" metric ($F(1, 31) = 10.77$, $p < 0.05$). Specifically, participants noticed the major events at a rate of 64.3% without the adaptive DST in the VA condition, compared to a significantly lower rate of 45.8% in the V condition. The higher rate of faster event notice suggests that the multimodal guidance in the VA condition effectively enhanced initial event detection.

Interestingly, however, the frequency of DST triggers did not differ significantly between the two groups ($F(1, 31) = 0.005$, $p = 0.94$). The adaptive DST was set to trigger not only when the participant overlooked the event but also when the participant failed to take immediate action post-detection. These results imply that although the VA condition facilitated faster initial noticing, the adaptive DST did not prompt MCs to make faster decisions once the event was perceived. Lastly, the rate at which they shifted their focus to the event area following the adaptive DST trigger was 9.8 percentage points higher in the V condition. However, this difference did not reach statistical significance ($F(1, 31) = 0.71$, $p = 0.41$).

**Table 3.** ART ANOVA of visual attention compared across the V and VA conditions.

| | $df_{numerator}$ | $df_{denominator}$ | *F* | *P Value* |
|---|---|---|---|---|
| **Notice the Event without DST** | 1 | 31 | 10.7661 | 0.0026* |
| **DST Triggered** | 1 | 31 | 0.0055 | 0.9415 |
| **Shift Attention Post-DST Trigger** | 1 | 31 | 0.7106 | 0.4057 |

## 5. DISCUSSION

The present study was designed to address three primary research questions: (1) whether an eye-tracker-triggered multimodal adaptive decision support tool, which integrates both visual and auditory modalities, enhances MCs' performance relative to a visual-only guidance condition; (2) how MCs' visual attention differs across the two conditions; and (3) which specific gaze-dynamics indicators serve as reliable predictors of that performance. The study's findings can be divided into three-folds: (1) intervention effect demonstrates that multimodal support enhances mission outcomes, as the VA condition significantly improved overall mission performance (MOP) compared to the V condition; (2)

the attention–performance fold identifies predictors of successful performance that are independent of experimental condition and nonlinear which indicates inverted-U relationships, such that moderate levels of recurrence and determinism were associated with optimal performance, whereas both low and high extremes were less beneficial; (3) the mechanism and boundary fold clarify the processes underlying the observed VA benefit and delineate its limitations, i.e., the VA condition improved event detection but didn't lead to quicker decision-making.

### 5.1. Attention Dynamics

The stepwise regression results suggested that the relationship between MCs' performance and attentional control via eye movements is nonlinear. This means that there is an 'optimal zone' of performance. A higher performance score was not necessarily linked to being perfectly 'locked in' certain scanning patterns (i.e., high determinism and RR) or having a totally unpredictable gaze pattern (i.e., low determinism and RR). Instead, the analysis shows that effective scanning involves combining gazes in certain patterns with scanning more freely and broadly, depending on the situation. Such a result aligns with the Yerkes-Dodson curve of arousal and performance, which shows that human performance increases with heightened physical or mental arousal but only up to a certain point (Yerkes & Dodson, 1908). As the Yerkes-Dodson curve shows, gathering too much or too little information through certain scanning behavior can hinder performance. This is supported by one of the findings that the observed curvilinear (inverted-U) relationship for RR indicates that team performance, as measured by MOP, was maximized when MCs engaged in moderate levels of attentional revisiting. Specifically, returning to key AOIs at a frequency sufficient to maintain ongoing monitoring processes, but not so frequently that attentional resources became constrained to a limited set of states, was associated with optimal outcomes. Within the context of C5ISR attention dynamics, low RR values indicate insufficient revisiting of previously sampled information, potentially leading to under-monitoring of critical task-relevant cues. Conversely, high RR values reflect excessive repetition of gaze patterns, potentially limiting the sampling of novel or emergent information in the task environment. These findings support the interpretation that effective command attention is characterized by a balance between the stability provided by revisiting and the flexibility required for novel sampling, rather than by maximizing either process.

Similar to the RR, the observed inverted-U relationship for determinism (DET) suggests that optimal performance in C5ISR monitoring tasks is associated with moderate levels of scanpath regularity. This finding aligns with the notion that structured, yet adaptable, routines characterize effective command attention. In this context, low DET values correspond to less organized, more opportunistic scanning patterns, which can compromise prioritization and systematic cross-checking. Conversely, excessively high DET values reflect over-routinization, resulting in attentional patterns that are overly predictable and insufficiently responsive to dynamic information requirements. These results indicate that determinism serves as an index of strategic organization, where an intermediate degree of regularity facilitates stable monitoring and integration while avoiding the inflexibility that can arise under increased workload or when capturing salient information. The Yerkes-Dodson curve can also provide further grounds for why the participants performed significantly better during the VA condition. Even though the auditory cue was only played when the participants weren't paying attention to the right things or when they had not taken an appropriate action after detecting the major events, the cue itself may have heightened the participants' arousal level, putting them in a more optimal state to be more vigilant, and therefore, perform better.

Within the context of C5ISR attention dynamics, ENTR quantifies the diversity of recurrent sequence lengths, reflecting the capacity for flexible switching among multiple monitoring strategies as task demands evolve. Such flexibility aligns with the concept of adaptive control, where effective attention coordination enables timely reconfiguration of scanning behaviors to accommodate shifting operational priorities. Notably, the relevance of ENTR in this context is not indicative of randomness but of structured flexibility: the ability to adaptively vary attention routines while maintaining coherent monitoring processes. Additionally, the significant association between entropy (ENTR) and improved performance, as indicated by lower MOP values, suggests that increased complexity and heterogeneity in recurrent gaze structures are associated with better performance. This finding suggests that high-performing MCs may employ a broader set of attention-allocation routines, rather than relying on a single, uniform pattern. The positive relationship with entropy indicates that a richer, more varied repertoire of attentional shifts enables mission commanders to gather information more effectively in complex environments. In such complex and dynamic environments, a uniform, repetitive gaze pattern can be a sign of vision and cognitive tunneling, negatively impacting MCs' ability to assess and make sound decisions.

The observed significant linear effect for mean diagonal line length (L) indicates that increased persistence in repeated gaze trajectories is associated with reduced performance, as reflected by higher MOP values, within this task context. Within the framework of C5ISR attention dynamics, a higher L value corresponds to prolonged, uninterrupted sequences of gaze behavior, in which attention remains focused on a specific scanning pattern for extended durations prior to transitioning. Although such persistence may facilitate exploitation under stable conditions, in dynamic, time-sensitive command environments, it can limit adaptability by delaying shifts to other relevant areas of interest (AOIs) in response to emerging cues. These results distinguish between a beneficial organizational structure, characterized by moderate revisiting and predictability, and excessive persistence, which may constrain the flexibility required for optimal team performance in complex socio-technical systems. This observation is in line with Endsley's description of demons of SA. One of the demons of SA is tunneling. Vision or cognitive tunneling occurs when an operator experiences cognitive overload and becomes locked into just a few pieces of information that may not be the most critical at that moment (Endsley et al., 2003). This results in failure to grasp the broader perspective during an operation and decreases the overall SA. The overall result shows the importance of adapting gaze behavior to the critical level of the situation. A similar result was found in a study conducted by Bammel (2003) exploring gaze behavior in the reading context. The study investigated gaze behaviors among beginners and experts in biology and physics. Interestingly, the expert group demonstrated more irregular eye movement dynamics while reading scientific texts. Bammel attributed irregular eye movement dynamics of experts to enhanced adaptivity in the reader-text coupling mechanism, which appears to facilitate the comprehension of the reading materials (Bammel, 2023).

Another interesting finding is the interaction between trial number and the curvilinear nature of RQA metrics, specifically, determinism ($DET^2$) and entropy ($ENTR^2$). These findings suggest that the influence of interaction structure on performance is not static but evolves with experience, reflecting both nonlinear and experience-dependent changes in the underlying dynamics. This means that when participants faced more novel situations in which they did not yet have effective strategies to counter events, the mission's success likely depended on being flexible in their gaze behavior and on exploring more widely to gather information and make the best decision possible. This finding aligns with a study that investigated the visual behavior of sailors at various ranks (top 10 and bottom 10 ranking). The study showed that higher-ranked sailors employed a more active visual search and more complex eye positions, which were associated with a better visual perception strategy. On the other hand, younger, lower-ranked sailors displayed more consistent and predictable visual behaviors (Manzanares et al., 2017; Menayo et al., 2018). These results provide an interesting contrast to similar studies on procedural tasks such as laparoscopic surgery. In a study involving laparoscopic procedures, highly structured and systemic gaze behavior was considered a sign of expertise and efficient visual attention (Zheng et al., 2021). However, in open-ended, dynamic environments such as command and control, complex gaze behaviors can be positively associated with mission outcomes.

### 5.2. Detection Versus Decision

According to Wickens' N-SEEV model, the probability of noticing an emerging event increases with event saliency and eccentricity (C. Wickens, 2015). The VA condition in this study benefited from higher saliency and eccentricity, as the event notice rate without the adaptive DST was significantly higher than in the V condition. The auditory cue also reduced the effort required for detection, which could explain why participants exhibited a lower attention shift rate during the VA condition. However, the increased notice rate did not necessarily prompt participants to make decisions regarding the events they observed. This is evident from the nonsignificant difference in DST trigger rate between the two conditions. The multimodal cues positioned the participants to be more sensitive to bottom-up saliency but did not override their internal expectancy or value regarding the urgency of the response. Therefore, further studies are necessary to explore methods not only to help MCs notice critical events but also to help them make necessary decisions after observing them. Nudging mission commanders to make decisions after noticing an event can be done not only by notifying them of the event, but also by providing possible COAs they can take in response.

### 5.3. Practical Implications for Real World Operations and Training

The results of this study can have practical implications for real-world operations and training of the next generation of mission commanders. Unlike procedural tasks, MCs are constantly operating in complex, dynamic environments. In these environments, perceiving the right information at the right time could greatly influence the decision quality. To better adapt to complex operational environments, training should teach future mission commanders to conduct rapid strategic scans of display sections to gather the information needed to make sound decisions for friendly forces. In the real world, the situations that MCs face are not always repeated. Therefore, even though they can draw on past experience, MCs should always be prepared to make adaptive and complex assessments during operations. Additionally, the multimodal (VA) systems can serve as training wheels during mission commanders' first exposure to complex environments, helping them stabilize their attentional control more quickly.

### 5.4. Limitations

This study had some notable limitations. First, the participants in this study, despite their military training, were not mission commanders, and the results may not generalize to MCs. More work is warranted to validate these findings with MCs. Second, attention may be impacted by arousal. In our study, we neither measured nor controlled for arousal levels. Finally, this study used only one eye-tracker, which primarily tracked gaze on the main map display. Future studies should employ several eye-trackers to investigate visual attention to several C5ISR interfaces.

## 6. CONCLUSIONS AND FUTURE WORK

Building on our prior work showing the efficacy of visual-only attention guidance for C5ISR tasks, this study explored the efficacy of multimodal adaptive attention guidance in further enhancing the mission commander's decision-making performance. Stepwise regression with BIC showed that utilizing multimodal adaptive attention cues can enhance performance compared with visual cues alone. Furthermore, having auditory cues in addition to visual cues put mission commanders in a more cognitively heightened state, enabling the MCs to detect major events at a significantly higher rate, even without adaptive attention guidance. However, faster event detection did not lead to faster decision-making. Therefore, a more robust and helpful adaptive decision-support tool should not only inform the MCs of events but also recommend COAs for ongoing events to expedite sound decision-making.

In future research, an additional study could investigate whether adding the recommended COA can enhance MCs' performance and reduce the cognitive and visual load imposed on mission commanders. Future research involving the recommended COAs could be in the context of human-AI teaming, where the COAs are formulated by an AI agent working as a teammate to human operators. This topic is highly relevant to contemporary technological advancements and the rapidly evolving battlefield.

## 7. ACKNOWLEDGEMENT

The authors thank [redacted for blind review], for reviewing the language and clarity of an earlier version of this manuscript. Additionally, the authors thank the many undergraduate and graduate students [redacted for blind review] who have provided tremendous help in executing the experiment and data analysis.